\documentclass{aa}
\usepackage{graphicx}
\usepackage{natbib}
\bibpunct{(}{)}{;}{a}{}{,} 
\def\ergs{ergs s$^{-1}$}
\def\ergscm{ergs s$^{-1}$ cm$^{-2}$}

\begin{document}

\title{Discovery of a quiescent neutron star binary in the globular cluster M13}

\author{Bruce Gendre\inst{1}, Didier Barret\inst{1}, Natalie
  Webb\inst{1}} \offprints{D. Barret (Didier.Barret@cesr.fr)}

\institute{Centre d'Etude Spatiale des Rayonnements, 9 Av. du Colonel Roche, 31028 Toulouse Cedex 04, France}
\date{Received / Accepted}

\abstract{We have discovered with XMM-Newton an X-ray source in the
  core of the globular cluster M13, whose X-ray spectral properties
  suggest that it is a quiescent neutron star X-ray binary.  The
  spectrum can be well fitted with a pure hydrogen atmosphere model,
  with T$_{\infty}=76\pm3$ eV, R$_{\infty}=12.8\pm0.4$ km and an X-ray
  luminosity of $7.3\pm0.6\times10^{32}$ \ergs. In the light of this
  result, we have discovered a strong correlation between the stellar
  encounter rate and the number of quiescent neutron stars found in
  the ten globular clusters observed so far by either XMM-Newton or
  Chandra. This result lends strong support to the idea that these
  systems are primarily produced by stellar encounters in the core of
  globular clusters.   } \titlerunning{XMM-Newton observation of M13}
\authorrunning{Gendre, Barret, Webb} \maketitle

\section{Introduction}

Globular clusters (GCs) are known to harbour both bright and dim X-ray
sources. Bright X-ray sources with luminosities above $\sim
10^{36}$\ergs\ are commonly agreed to be neutron star Low-Mass X-ray
Binaries. There are twelve such sources known in GCs; 11 of them have
shown type I X-ray bursts, the unmistakable signature of an accreting
neutron star. On the other hand, dim GC X-ray sources are much more
numerous \citep[for example, dozens of such objects exist in
$\omega$~Cen, see][]{gen03}, but their nature is still debated. They
have maximum luminosities of $\sim 10^{33}$ \ergs\ and there is
growing evidence that they are a variety of different objects. Some of
the fainter ones have been identified as Cataclysmic Variables or
active binaries (RS CVn, BY Dra). Others, have been associated with
radio millisecond pulsars (MSPs) \citep[e.g.][]{gri91}. Some of the
`brighter' of the dim X-ray sources have recently been proposed to be
quiescent neutron star X-ray binaries (qNSs) on the basis of their
X-ray spectral properties (47 Tuc, Grindlay et al.  2001; $\omega$
Cen, Rutledge et al.  2002, Gendre et al.  2002; NGC 6440, Pooley et
al. 2002b).

Outside GCs, qNSs are found to have soft X-ray spectra and
luminosities up to $10^{33}$ \ergs\ \citep[e.g. ][]{kon02}. During
outbursts, reactions, deep in the crust, heat the neutron star
\citep{bro98}.  Between outbursts, the heated surface radiates a
thermal spectrum, emitted by a neutron star hydrogen atmosphere (NSA).
The thermal component of the quiescent X-ray spectra of two neutron
star X-ray binaries which are not in a globular cluster (Aql X-1 and
Cen X-4) are well fitted by NSA models, providing strong support for
this theory. In contrast, quiescent black hole transients have hard
power law like spectra and low luminosities (down to $10^{30}$ \ergs).
Accretion via an advection dominated flow is thought to be responsible
for the observed hard and weak emission \citep[][and references
therein]{kon02,ham02}.  Thus the combination of a soft thermal X-ray
spectrum and a luminosity above $\sim 10^{32}$ \ergs~has been used to claim
the detection of qNSs in GCs. The existence of such systems in GCs was
first proposed by \cite{ver84}. It is generally assumed that they are
formed through stellar encounters (tidal capture or exchange
encounters) in the dense cores of GCs \citep[see][ for a
review]{hut92}.

ROSAT HRI and PSPC observations \citep{fox96,ver01}, revealed several
dim X-ray sources in M13. The HRI observations showed two core
sources, whose unresolved PSPC spectrum could be fitted with a 0.9 keV
thermal bremsstrahlung model (L$_{0.5-2.5 \rm keV} \sim
2.5\times10^{32}$ \ergs). M13 is somewhat remarkable as it contains
also four MSPs \citep{tay93,ran02}.
\section{Observations and results}

The globular cluster M13 was observed by the XMM-Newton EPIC cameras
on 2002 January 28 and 30, using the Full Frame Window and a medium
filter. The total length of the observation was $\sim 37$ ks.  We
analyzed the data using the XMM-Newton Science Analysis Software (SAS)
version 5.3.3. Initially we considered the two segments of the
observation separately. We used the SAS tasks {\it emchain} and {\it
  epchain} to calibrate the raw data, flag bad pixels and filter for
non-astrophysical events. The background was found to be variable and
relatively high.  Removing the periods of unstable background left 9
and 8 ks respectively. The filtered event files were merged using {\it
  merge} and then we extracted images and spectra.

The source detection procedure used is described in \cite{gen03}.
Briefly, it combines a wavelet detection algorithm applied to a 0.5 -
5.0 keV band image, and a maximum likelihood fitting of the source
candidates. A conservative maximum likelihood threshold of 12 was
chosen. The absorption-corrected limiting flux is $3.6 \times
10^{-15}$ \ergscm\ corresponding to a luminosity of $2.6 \times
10^{31}$ \ergs~for a source located at the center of the field of view
having a 0.6 keV blackbody spectrum. 

\begin{figure}[!t]

\centerline{\includegraphics[width=9.5cm]{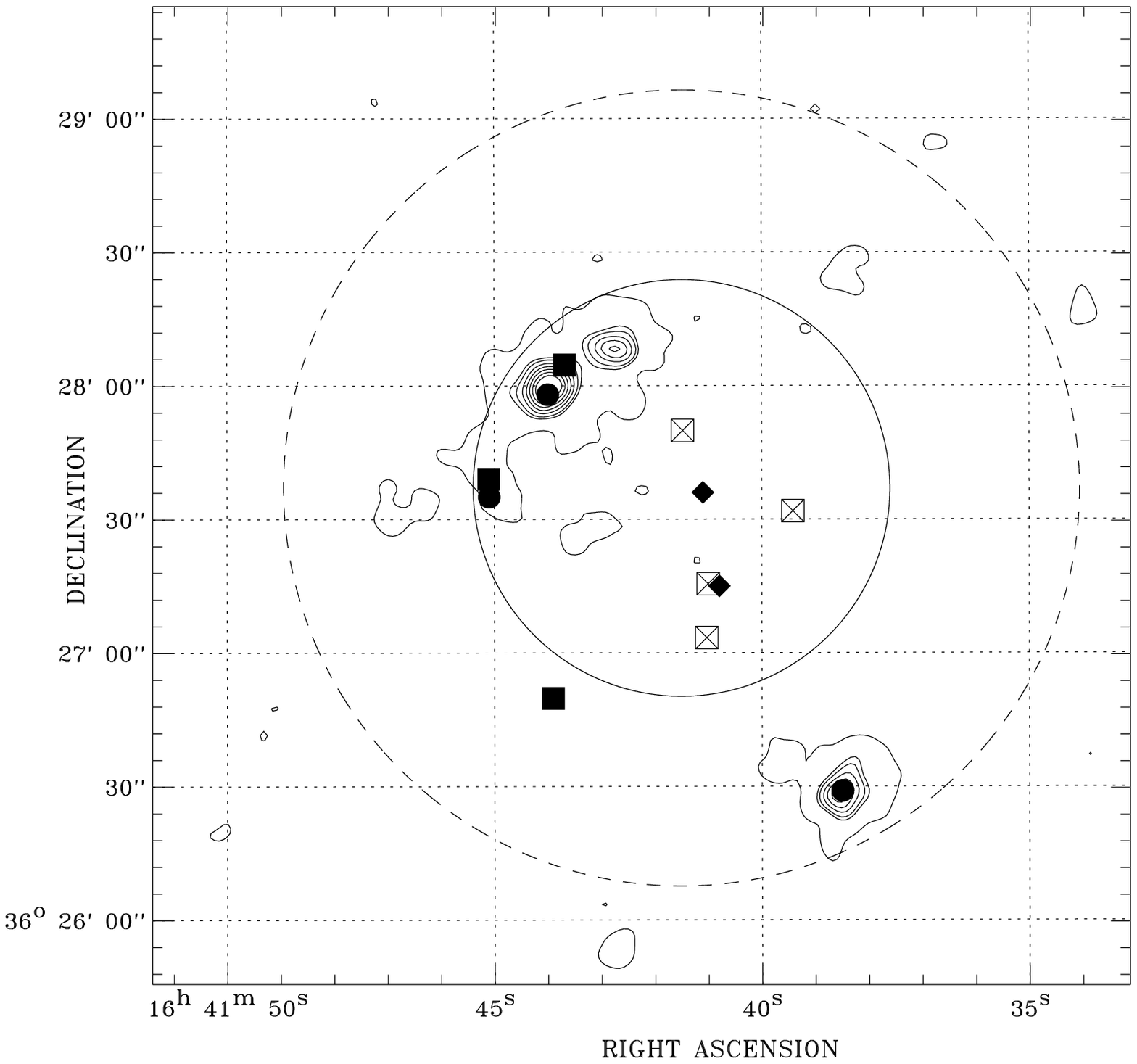}}

\caption[]{A contour image of the center of the field of view of M~13. The core and half mass radii are shown as solid and dashed lines, respectively. Previous identifications are: ROSAT sources \citep[filled circles, ][]{ver01}; faint UV sources \citep[filled squares, ][]{fer97}; radio objects \citep[open squares,][]{joh91}; and MSPs with known positions \citep[filled diamonds,][]{tay93} (the most central MSP has positional uncertainties larger than the image displayed.  All other objects have uncertainties smaller than the symbol size).}

\label{gendre_f1}  

\end{figure}

We found an extended source within the core radius.  The extension
appeared to be due to two sources that were not quite resolved. The
MOS cameras have the advantage of having a smaller pixel size than the
PN camera: 1.1\arcsec\ for MOS \citep{tur01} as oppose to 4.1\arcsec\ 
for PN \citep{str01}. When processing the event file, the task {\it
  emevents} converts the event position (RAWX and RAWY) into camera
coordinates in units of 0.05\arcsec. This step includes randomization
(within a CCD pixel) of the event to avoid the Moir\'e effect. By
default, the {\it xmmselect} task, assumes a binning factor of 87 to
produce MOS images with the same pixel size as PN images. We have
produced MOS images with a binning size of 20, corresponding to a
pixel size of 1\arcsec. The image smoothed with a simple 2D-Gaussian
of $\sigma$ = 2.0\arcsec\ is shown in Fig \ref{gendre_f1}.

To summarize, we detected 2, 5 and 77 sources within the core radius, half mass
radius and field of view respectively. Using the Log N--Log S curve of
extragalactic sources reported in \citet{has01} and the limiting flux
in the three regions given above, we determined the expected number of
background sources to be 0, 1 and 72, respectively. The positions and
errors of sources found within twice the half-mass radius are given in
Table~\ref{gendre_table1}.  It goes beyond the scope of this paper to
investigate the error box content of each of the XMM-Newton sources.
It is however worth mentioning that only 10 of the 11 ROSAT sources
that should have been detected by XMM-Newton, given the limiting
flux and field of view of the observation, were detected. We failed to
detect the ROSAT source Gb. This source must have therefore varied in
flux by at least a factor 10 between the ROSAT and XMM-Newton
observations.

\begin{table}[!thb]

\caption{XMM-Newton sources detected within twice the half mass radius (HMR) (CR = Core Radius). The positional error includes the statistical error (90\% confidence, estimated from {\it srcmatch}) and a systematic error of 4\arcsec \citep[][]{jan01}. 
\label{gendre_table1}}

\begin{center}

\begin{tabular}{ccccc}

\noalign{\smallskip\hrule\smallskip}
 R.A.   & Dec.   & Error & Location \\

  h  m  s  & $\degr$ $\arcmin$ $\arcsec$ & \arcsec  \\

\noalign{\smallskip\hrule\smallskip}

16 41 37.9 & 36 28 26.3 &   7.94 & HMR \\
16 41 42.7 & 36 28 06.7 &   4.60 & CR \\
16 41 43.8 & 36 27 58.6 &   4.27 & CR \\
16 41 46.8 & 36 27 29.5 &   5.93 & HMR \\
16 41 49.0 & 36 26 44.6 &   7.53 & 2$\times$HMR\\
16 41 38.3 & 36 26 27.0 &   4.50 & HMR \\
16 41 49.9 & 36 26 18.7 &   7.38 & 2$\times$HMR\\
16 41 42.5 & 36 25 53.0 &   6.83 & 2$\times$HMR\\
\noalign{\smallskip\hrule}
\end{tabular}
\end{center}
\end{table}

The two core sources are separated by only 15\arcsec (see Fig
\ref{gendre_f1}).  One is the ROSAT source Ga \citep{ver01}. The other
should have been detected by ROSAT, if its flux had not varied. We
reanalyzed the HRI data and determined that the source flux must have
varied by a factor $\sim 2$ between the ROSAT and the XMM-Newton
observations.

Two sources lying so close together complicates the spectral analysis.
Normally, spectra are accumulated over a region of radius of
0.7\arcmin~to include 85\% of the source photons.  Such an extraction
region is 3 times larger than the source separation. Using a radius of
0.7\arcmin, we have extracted the spectrum of the brighter source by
masking out a region of radius 15\arcsec, offset by 5\arcsec\ from the
fainter source. We have estimated that only 6\% of the counts in the
spectrum of the brighter source come from the fainter source,
insufficient to affect the results of our spectral analysis.  Spectra
were extracted from the EPIC-PN and the two MOS cameras. We binned
these spectra to contain at least 20 net counts per bin and generated
ARF and RMF files with the SAS tasks {\it arfgen} and {\it rmfgen}.
The spectra are extremely soft, with $\sim 90$\% of the counts below 2
keV. We have tried to fit the combined spectrum with different single
component models (blackbody and thermal bremsstrahlung).  Absorption by
the interstellar medium was included in the fit but was found to be
consistent with the expected value from the optical extinction towards
the cluster \citep[$1.1 \times 10^{20}$ cm$^{-2}$, ][]{djo93}.  We also tried to fit
the spectrum with a pure hydrogen NSA model \citep{pav92,zav96}.  This
model provides the best fit to our data.  Assuming a neutron star mass
of 1.4M$_\odot$, we derived T$_{\infty}=76^{\scriptscriptstyle
  +3}_{\scriptscriptstyle -3}$ eV and
R$_{\infty}=12.8^{\scriptscriptstyle +0.4}_{\scriptscriptstyle -0.4}$
km, with a $\chi^{\scriptscriptstyle 2}_{\scriptscriptstyle \nu}$=0.55
(15 degrees of freedom (dof)).  These parameters are similar to those
determined for the proposed qNS in $\omega$ Cen (Rutledge et al. 2002,
Gendre et al. 2003).

\begin{figure}[!tbh]  
\centerline{\includegraphics[width=9.5cm]{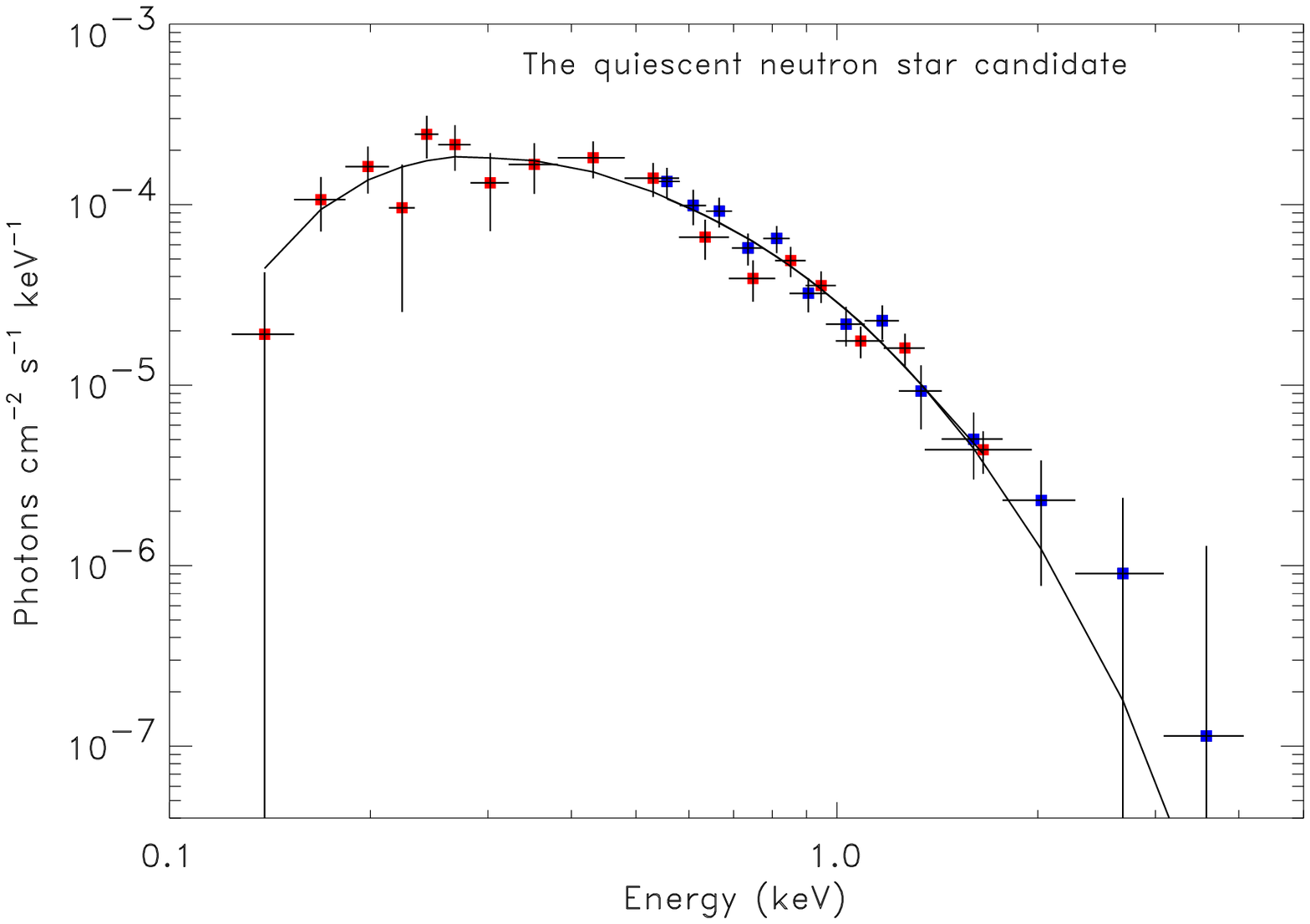}}

\caption[]{The unfolded EPIC-PN (blue points) and ROSAT-PSPC (red points) spectra of the qNS candidate and the best fit NSA model, one of the best qNS spectra to date.}

\label{gendre_f2}

\end{figure}

We have retrieved and reanalyzed the ROSAT PSPC archival observations
of M13 to determine whether the PSPC spectrum of source Ga could
be fitted with the same model. A fit of the combined PSPC and XMM-Newton
data revealed parameters consistent with those derived from the
XMM-Newton data alone ($\chi^{2}_{\nu} = 0.90$, 26 dof).  We
present the unfolded combined EPIC-PN and ROSAT spectrum in Fig.
\ref{gendre_f2}. The luminosity derived for the NSA model is
($7.3 \pm 0.6) \times 10^{32}$ erg s$^{-1}$ (0.1-5.0 keV), using
the distance of 7.7 kpc (Harris 1999).

\section{Discussion}
The most likely interpretation of the nature of the softest source in
M13 is that it is a quiescent neutron star low-mass X-ray binary. This
interpretation is supported by the luminosity of the source, the
softness of its X-ray spectrum and the fact that a NSA model yields a
good fit and plausible parameters (radius and temperature) for the
neutron star.  A reasonable question to ask is whether we expect such
a system in M13. In the disk, qNSs have a minimum X-ray luminosity of
$\sim 10^{32}$ \ergs \citep{nar02}. As it has been already emphasized, if
the same luminosity threshold also applies to globular cluster qNSs,
the luminosity limit of XMM-Newton and Chandra observations (typically
around $10^{30}-10^{31}$ \ergs\ at the cluster distances) allows one to
detect all the qNSs present in globular clusters. In table
\ref{gendre_table3}, we list the globular clusters already observed by
either Chandra or XMM-Newton, together with the number of qNSs
reported in the literature.

\begin{figure}[!t]  
\centerline{\includegraphics[width=8.0cm]{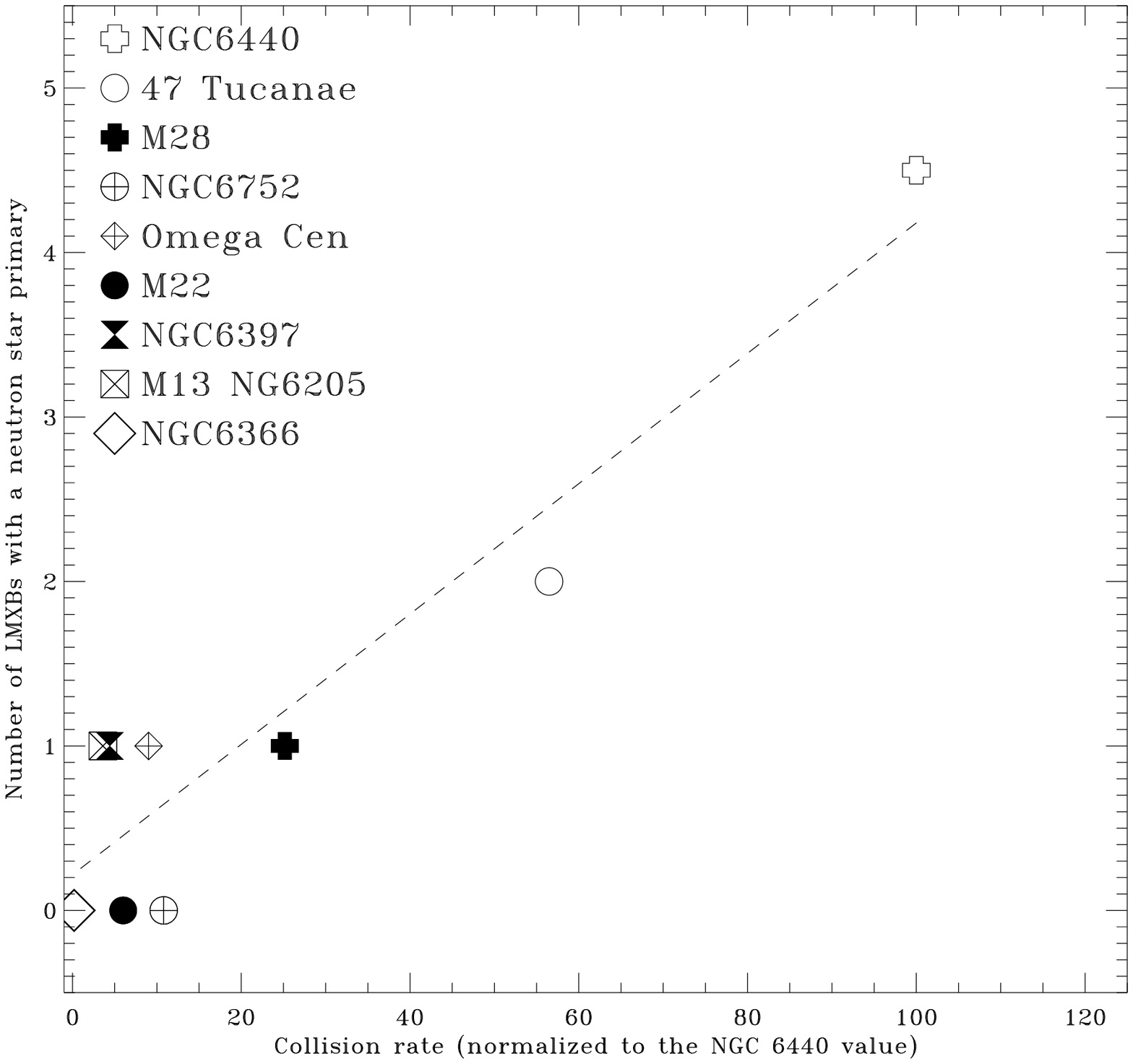}}
\caption[]{The number of qNSs versus collision rate ($\rho_0^{1.5} r_c^2$), shown with a linear fit $n_{\rm qNS} \sim 0.04 \times \rho_0^{1.5} r_c^2 + 0.2$. The collision rates have been normalized to 100 for NGC 6440.} 
\label{gendre_f3}

\end{figure}

In globular clusters, the number of qNSs is expected to scale with the
collision rate which is proportional to $\rho_0^{1.5} r_c^2$ for
virialized clusters, where $\rho_0$ is the central density of the
cluster and $r_c$ its core radius \citep{ver02}. These values
taken from the \citet{har99} catalog are listed in Table
\ref{gendre_table3}. In Fig \ref{gendre_f3}, we plot the number of
qNSs as a function of the collision rate, normalized so that the value
for NGC~6440 is 100. There is a striking correlation between the number
of qNSs and the collision rate. The presence of one qNS in M13 is
therefore not really surprising given that its collision rate puts the
cluster in a region where one might expect either one or zero qNS.
This remarkable correlation extends over more than 2 orders of
magnitude and includes both core-collapsed and non core-collapsed
clusters. This strongly supports the idea that qNSs are indeed
primarily produced by stellar encounters in globular clusters
\citep{ver02}.

With the results of the observations reported here and the four
already known MSPs, there are at least 5 neutron star
systems in M13. This makes M13 the cluster with the fourth highest
number of known neutron star systems. The retention of such a large
number of neutron stars in a cluster with a relatively low central
density remains to be explained \cite[see for a comprehensive study of
neutron star retention in globular clusters,][]{pfa02} .

\begin{table}[t!]
\caption{qNSs in GCs detected by Chandra or XMM-Newton. Parameters for the clusters are taken from the \citet{har99} catalog; the distance is given in kpc, the core radius in \arcmin. We indicate the log of the central density in units of L$_\odot pc^{-3}$.\label{gendre_table3}}
\begin{center}
\begin{tabular}{lccccc}
\noalign{\smallskip\hrule\smallskip}
Cluster & Distance &Core   & Central & qNSs \\
   & (kpc)       &radius & density &       \\
\noalign{\smallskip\hrule\smallskip}
47 Tuc       & 4.5 & 0.44 & 4.77 & 2  \\
$\omega$ Cen & 5.3 & 2.58 & 3.12 & 1 \\
M 13         & 7.7 & 0.78 & 3.33 & 1        \\
NGC 6366     & 3.6 & 1.83 & 2.42 & 0   \\
NGC 6397     & 2.3 & 0.05 & 5.68 & 1  \\
NGC 6440     & 8.4 & 0.13 & 5.28 & 4-5     \\
M 28         & 5.7 & 0.24 & 4.75 & 1        \\
M 22         & 3.2 & 1.42 & 3.64 & 0        \\
NGC 6752     & 4.0 & 0.17 & 4.91 & 0       \\
\noalign{\smallskip\hrule\smallskip}
\end{tabular}
\end{center}
References, top to bottom: \citet{gri01,gri02}; \citet{rut02}, \citet{gen03}; this work; Webb, Barret, Gendre, in preparation; \citet{gri01b}; \citet{poo02b};  \citet{bec03};  \citet{web02}; and \citet{poo02a}.
\end{table}

\section{Conclusion}

We have reported the likely discovery of a quiescent neutron star in
M13. We have also shown for the first time that there is a strong
correlation between the stellar collision rate and the number of
qNSs in the 9 globular clusters observed by either XMM-Newton or
Chandra. This lends strong support to the idea that these systems are primarily
produced by stellar encounters in the core of globular clusters. More observations, in particular with XMM-Newton are
being planned and should enable us to test the strength of this
correlation.

\begin{acknowledgements}
  
  We wish to thank the referee, Slava Zavlin, for his comments on this
  manuscript and Franck Verbunt and Craig Heinke for their additional remarks.

\end{acknowledgements}


\begin{thebibliography}{}

\bibitem[Becker et al. (2003)]{bec03} Becker, W., Swartz, D., Pavlov, G., et al., 2003, \apj, in press (astro-ph 0211468)
\bibitem[Brown et al. (1998)]{bro98} Brown, E.F., Bildstein, L., \& Rutledge, R.E., 1998, \apj, 504, L95
\bibitem[Djorgovski (1993)]{djo93} Djorgovski, S. G. 1993, \pasp, 50, 373
\bibitem[Fox et al. (1996)]{fox96} Fox, D., Lewin, W., Margon, B., van Paradijs, J., \& Verbunt, F., 1996, \mnras, 282, 1027
\bibitem[Ferraro et al. (1997)]{fer97} Ferraro, F. R., Paltrinieri, B., Fusi Pecci, F., Rood, R. T., Dorman, B., 1997, \mnras, 292, L45
\bibitem[Gendre et al. (2003)]{gen03} Gendre, B., Barret, D., \& Webb, N. A., 2003, \aap, 400, 521
\bibitem[Grindlay et al. (1991)]{gri91} Grindlay, J. E., Cool, A. M., Baylin, C. D., 1991, in The formation and evolution of star clusters, ASP Conf. Serie Vol 13, 396
\bibitem[Grindlay et al. (2001a)]{gri01} Grindlay, J. E., Heinke, C., Edmonds, P. D., \& Murray, S. S., 2001, Sci., 292, 2290
\bibitem[Grindlay et al. (2001b)]{gri01b} Grindlay, J. E., Heinke, C., Edmonds, P. D., Murray, S. S., \& Cool A. M., 2001, \apj, 563, L63
\bibitem[Grindlay et al. (2002)]{gri02} Grindlay, J.E., Camilo, F., Heinke, C., et al. 2002, \apj, 581, 470
\bibitem[Hameury et al. (2003)]{ham02} Hameury, J.M., Barret, D., Lasota, J.E., et al. 2003, \aap, 391, 631
\bibitem[Harris (1999)]{har99} Harris, W. E. 1996, Ap\&SS, 267, 95
\bibitem[Hasinger et al. (2001)]{has01} Hasinger, G., Altieri, B., Arnaud, M., et al. 2001, \aap, 365, L45
\bibitem[Hut et~al. (1992)]{hut92} Hut, P., McMillan, S., Goodman, J., et al. 1992, \pasp, 104, 981
\bibitem[Jansen et~al. (2001)]{jan01} Jansen, F., Lumb, D., Altieri, B., et al. 2001, \aap, 365, L1
\bibitem[Johnston et al. (1991)]{joh91} Johnston, H. M., Kulkarni, S. R., \& Goss, W. M., 1991, \apj, 382, L89
\bibitem[Kong et al. (2002)]{kon02} Kong, A. K. H., McClintock, J. E., Garcia, M. R., Murray, S. S., \& Barret, D. 2002, \apj, 570, 277
\bibitem[Narayan et al. (2002)]{nar02} Narayan, R., Garcia, M.R., \& McClintock, J.E. 2002, in Proc. IX Marcel Grossmann Meeting, eds. V. Gurzadyan, R. Jantzen and R. Ruffini, Singapore : World Scientific, in press, astro-ph/0107387
\bibitem[Pavlov et al. (1992)]{pav92} Pavlov, G. G., Shibanov, Y. A., \& Zavlin, V. E. 1992, \mnras, 253, 193 
\bibitem[Pfahl et al. (2002)]{pfa02} Pfahl, E., Rappaport, S., \& Podsiadlowski, P. 2002, \apj, 573, 283
\bibitem[Pooley et~al. (2002a)]{poo02a} Pooley, D., Lewin, W. H. G., Homer, L., et al. 2002a, \apj, 569, 405
\bibitem[Pooley et~al. (2002b)]{poo02b} Pooley, D., Lewin, W. H. G., Homer, L., et al. 2002b, \apj, 573, 184
\bibitem[Ransom et al. (2002)]{ran02} Ransom, S. M., Hessels, J. W. T., Stairs, I. H., et al, 2002, To appear in "Radio Pulsars" (ASP Conf. Ser.), eds M. Bailes, D. Nice, \& S. Thorsett (astro-ph 0211160)
\bibitem[Rutledge et al. (2002)]{rut02} Rutledge, R. E., Bildsten, L., Brown, E. F., Pavlov, G. G., \& Zavlin, V. E. 2002, \apj, 578, 405
\bibitem[Str\"{u}der et~al. (2001)]{str01} Str\"{u}der, L., Briel, U., Dennerl, K., et al. 2001, \aap, 365, L18
\bibitem[Taylor et al. (1993)]{tay93} Taylor, J. H., Manchester, R. N., \& Lyne, A. G., 1993 \apjs, 88, 529
\bibitem[Turner et al. (2001)]{tur01} Turner, M. J. L., Abbey, A., Arnaud, M., et al. 2001, \aap, 365, L27
\bibitem[Verbunt et al. (1984)]{ver84} Verbunt, F., Elson, R., \& van Paradijs, J., 1984, \mnras, 210, 899
\bibitem[Verbunt (2001)]{ver01} Verbunt, F. 2001, \aap, 368, 137
\bibitem[Verbunt (2002)]{ver02} Verbunt, F. 2002, to appear in ASP Conf. Ser., New horizons in globular cluster astronomy, ed. G. Piotto, G. Meylan, G. Djorgovski \& M. Riello (astro-ph 0210057)
\bibitem[Webb et~al. (2002)]{web02} Webb, N. A., Gendre, B., \& Barret, D. 2002, \aap, 381, 481
\bibitem[Zavlin et al. (1996)]{zav96} Zavlin, V. E., Pavlov, G. G., \& Shibanov, Y. A. 1996, \aap, 315, 141 
\end{thebibliography}
\end{document}